\documentclass[%
 reprint,
 amsmath,amssymb,
 aps,
prb,
]{revtex4-2}

\usepackage{graphicx}% Include figure files
\usepackage{dcolumn}% Align table columns on decimal point
\usepackage{bm}% bold math
\begin{document}

%\preprint{APS/123-QED}

\title{DFT Investigation of Magnetocrystalline Anisotropy \\ in Fe, Co, Pd$_{0.97}$Co$_{0.03}$ and Pd$_{0.97}$Fe$_{0.03}$ systems: From Bulk to Thin-Films}% Force line breaks with \\
%\thanks{A footnote to the article title}%

\author{Irina I. Piyanzina}
 %\altaffiliation[Also at ]{Physics Department, XYZ University.}%Lines break automatically or can be forced with \\
\email{i.piyanzina@gmail.com}
\author{Hayk Zakaryan}
\affiliation{Center of Semiconductor Devices and Nanotechnology, Computational Materials Science Laboratory, Yerevan State University, Republic of Armenia}%

%\collaboration{MUSO Collaboration}%\noaffiliation
\author{Regina M. Burganova}
\author{Zarina I. Minnegulova}
\author{Igor V. Yanilkin}
\author{Amir I. Gumarov}
\altaffiliation[Also at ]{E.K. Zavoisky Physical-Technical Institute of RAS, Russia}%
% \homepage{http://www.Second.institution.edu/~Charlie.Author}
\affiliation{Institute of Physics, Kazan Federal University, Russia}%

\date{\today}% It is always \today, today,
             %  but any date may be explicitly specified

\begin{abstract}
The nature of low-impurity ferromagnetism remains a challenging problem in the solid-state community due to the strong dependence of magnetic properties on composition, concentration, and structural geometry of diluted alloys. To address this, we performed a density functional theory study of magnetocrystalline anisotropy in Fe, Co, Pd$_{0.97}$Co$_{0.03}$, and Pd$_{0.97}$Fe$_{0.03}$ systems across bulk, monolayer, and thin-film geometries. Non-collinear spin-orbit calculations were employed to evaluate the magnetocrystalline anisotropy energies, supported by analysis of atomic-, spin-, and orbital-resolved densities of states. The results revealed that Fe and Co exhibit opposite easy-axis orientation depending on geometry. At the same time, even 3\% Co-doping in Pd is sufficient to induce anisotropy trends resembling those of pure Co. In contrast, Pd–Fe systems at the same concentration do not reproduce the anisotropy of pure Fe, showing isotropic behavior in bulk. 

%\begin{description}
%\item[Usage]
%Secondary publications and information retrieval purposes.
%\item[Structure]
%You may use the \texttt{description} environment to structure your abstract;
%use the optional argument of the \verb+\item+ command to give the category of each item. 
%\end{description}
\end{abstract}

%Use showkeys class option if keyword
                              %display desired
\maketitle

%\tableofcontents

\section{Introduction}
\label{intro}
Palladium-based thin films doped with magnetic impurities play an important role in the design of advanced magnetic devices, such as magnetic storage systems, sensors, etc., due to their unique magnetic properties~\cite{yanilkin2025temperature,Rakesh2022Magnetic}. These thin films demonstrate enhanced magnetic behavior, such as low coercive fields and tunable magnetization at low impurity concentrations~\cite{Jussila}. Understanding how ferromagnetic impurities influence the magnetic properties of binary palladium (Pd) alloys is essential for advancing materials science. 

Recently, "superconductor-ferromagnet-superconductor" structures have garnered attention due to their potential use in superconducting spintronics~\cite{arham,bolginov,larkin,ryazanov2,vernik,niedziel,glick,soloviev,usp}. In particular, Pd-Fe alloys hold promise for Josephson-junction-based superconducting magnetic random access memory (MRAM)~\cite{larkin,ryazanov2,soloviev,golovchanskiy}. Pd-rich ferromagnetic compositions with impurity concentrations between $0.01 < x < 0.1$ are particularly significant due to their unique magnetic characteristics. These alloys are generally produced as thin films, often through methods such as magnetron sputtering~\cite{arham,bolginov,larkin,ryazanov2,vernik,niedziel,glick,usp,uspenskaya1,uspenskaya2}, molecular beam epitaxy (MBE)~\cite{uspenskaya2,garifullin,ewerlin,esmaeili,esmaeili2,petrov,esmaeili3,mohammed,ya}, and ion implantation~\cite{gumarov,claus1986effect,hitzfeld1983ferromagnetism,hitzfeld1984ferromagnetism}. Experimental studies have shown that introducing Fe or Co into Pd enhances coercivity and saturation magnetization, while Ni and Mn reduce coercivity, which is advantageous for MRAM applications requiring lower switching currents~\cite{esmaeili,esmaeili2,petrov,esmaeili3,mohammed,ya,gumarov,gumarov2}. Thus, selecting the optimal impurity type and concentration requires careful balancing of the magnetic and structural properties.

Magnetic anisotropy -- the directional dependence of magnetic behavior -- plays a key role in magnetic systems. Magnetocrystalline anisotropy (MCA), caused by spin-orbit coupling and crystal symmetry, defines preferred magnetization directions. For example, cobalt's easy axis in hexagonal closed-packed (HCP) structures aligns along the [0001] direction~\cite{bulk_co}. MCA, intrinsic to the material, is crucial for applications like data storage (MRAM), permanent magnets, and sensors. For example, high anisotropy in materials, such as samarium cobalt (SmCo), allows the production of strong permanent magnets that retain their magnetization even at high temperatures~\cite{Gutfleisch2009}.
Shape anisotropy, in contrast, originates from dipole-dipole interactions and depends on the sample's geometry. In magnetic thin films, magnetization aligns more easily along the plane due to demagnetizing fields~\cite{KIRCHMAYR20014754}. 

In Pd-based alloys, magnetic anisotropy is closely tied to chemical ordering. For instance, research by P.~Kamp et al.~\cite{kamp99} on Pd-Fe with 50\,at.\% of Fe has shown perpendicular anisotropy in the ordered L$1_0$ phase (characterized by alternating layers of two types of atoms along a specific crystallographic axis). In contrast, the disordered $\gamma$ phase (when impurity atoms randomly replace atoms of the host matrix) demonstrates in-plane anisotropy to minimize demagnetizing energy. 
 
At the same time, monocrystals of Pd-Fe possess an easy axis along the [111] direction~\cite{bagguley1974resonance}. However, as reported by Esmaeli and co-workers, in Pd-Fe films it shifts to [110] due to shape anisotropy, resulting in four-fold in-plane magnetic anisotropy~\cite{esmaeili3}. Similarly, in V/Pd-Fe bilayers, FMR studies revealed the presence of four-fold in-plane MCA in thin films oriented toward the [001] direction, with easy axes along specific crystallographic directions~\cite{garifullin2007transformation}.  

Ganshina and coauthors reported the presence of strong perpendicular magnetic anisotropy in Pd-Co alloy films with 20\% of Co, highlighting their potential application as magneto-optical recording media~\cite{ganshina}. This anisotropy is attributed to thermal tensile stresses and significant negative magnetostriction, which induce a stress-driven perpendicular easy-axis anisotropy that overcomes the shape anisotropy.
 
Despite these advancements, the atomic-scale mechanisms underlying magnetic anisotropy in Pd-based alloys remain poorly understood. Systematic studies, particularly those involving \textit{ab initio} calculations, are needed to clarify how composition, impurity concentration, and structural geometry influence the magnetic anisotropy energy (MAE). Limited \textit{ab initio} calculations of Pd- and Pt-based alloys, studying MCA, exist. For example, the Density Functional Theory (DFT) study of Fe-Pd alloys across different phases has shown that chemical ordering strongly influences MCA~\cite{PATHAK2020166266}. The transition from a disordered phase to an ordered one induces a shift from in-plane to perpendicular anisotropy, attributed to the emergence of uniaxial MCA that dominates over demagnetizing effects. Furthermore, studies indicate the presence of intermediate tetragonal and orthorhombic phases in Fe-Pd, affecting coercivity and Curie temperature. 

DFT calculations on L$1_0$-ordered bulk Fe-\textit{M} alloys (\textit{M} = Ni, Pd, Pt) confirmed their ferromagnetic stability with magnetization out of the plane~\cite{aledealat2021first}. The highest MCA was found for FePt, while FePd and FeNi showed lower values, emphasizing MCA's sensitivity to structure and composition.

Beyond Pd- and Pt-based alloys, DFT has been extensively applied to magnetic anisotropy features calculations in transition-metal thin films and oxides. In Fe/W (110) and Fe/Mo (110) films, interface effects lead to significant MCA variations, with Fe/W exhibiting a strong in-plane anisotropy due to hybridization effects at the interface~\cite{qian2001}.
For transition metal monoxides (MnO, FeO, CoO, NiO), DFT+\textit{U} calculations have been used to capture the role of electron-electron interactions affecting MCA. The results show that spin-orbit coupling dominates MCA in FeO and CoO due to partially filled $t_{2g}$ orbitals, while MnO and NiO exhibit strain-induced anisotropy effects~\cite{Schr115134}. Additionally, DFT studies on Fe, Co, and Ni slabs have revealed that MCA oscillates with film thickness, highlighting the role of quantum confinement and surface effects~\cite{Laurent174426}.

Recently, magnetic skyrmions, relevant to spintronics, have been observed in magnetic thin films. In Co films, sub-10\,nm skyrmions can be stabilized at zero field by exchange frustration, Dzyaloshinskii–Moriya interaction (DMI), and large magnetocrystalline anisotropy~\cite{meyer2019isolated}. Further, DFT studies~\cite{nickel2023exchange} revealed that while exchange frustration weakens in Rh/Co/Ir multilayers with thicker Co layers, introducing Fe restores strong frustration and DMI, stabilizing skyrmions in Rh/Co/Fe/Ir systems.

Altogether, the presented literature overview revealed that magnetic properties depend significantly on the composition, geometry, film thickness, impurity type, its distribution, and concentration.
This study aims to systematically investigate magnetic properties and anisotropy of bulk, monolayers, and thin film Fe and Co, and to apply the developed methodology to diluted Pd-Co alloy in disordered $\gamma$-phase in the bulk and thin-film geometries using non-collinear spin–orbit DFT calculations. This work extends our previous research on magnetic moment dependence on impurity concentration in Pd-based diluted alloys~\cite{piyanzina_crystals,piyanzina_surface,our_alloys_paper}. Together, these studies provide a comprehensive understanding of electronic structure and magnetic anisotropy features in low-impurity magnetic alloys.

\section{Computational Details}
\label{computations}
Our $\textit{ab initio}$ investigations were based on the DFT~\cite{hohenberg1964,kohn1965} approach within the VASP code~\cite{kresse1996a,kresse1996b,kresse1999} as a part of the MedeA\textsuperscript{\textregistered} software of Materials Design~\cite{medea}. The effects of exchange and correlation were taken into account by the generalized gradient approximation (GGA) as parameterized by Perdew, Burke, and Ernzerhof (PBE)~\cite{perdew1996}. The Kohn-Sham equations were solved using the set of plane waves (PAW)~\cite{bloechl1994paw}. The cut-off energy was chosen to be equal to 400\,eV. The force tolerance was 0.5\,eV/nm, and the energy tolerance for the self-consistency loop was 10$^{-6}$\,eV. The Brillouin zones were sampled using Monkhorst--Pack grids~\cite{monkhorst1976}, including 9 $\times$ 9 $\times$ 1 {$\textbf{k}$-}points for monolayers and 15 $\times$ 15 $\times$ 15 for bulk cells of pure components. For the alloys 3 $\times$ 3 $\times$ 3 {$\textbf{k}$-}points grid was used. We performed spin-polarized calculations in all cases, initializing the impurity atoms to have initial magnetic moments and Pd atoms to be in the paramagnetic state. The electronic densities of states were calculated using the linear tetrahedron method~\cite{bloechl2} on the same {\bf k}-point grids with a 0.03 integration step.

The simplified GGA+\textit{U}~\cite{Dudarev} approach was applied to account for strong electron correlations. For the PAW potentials, with \textit{d}-electrons treated as valence states, we used \textit{U}$_d$ values of 3.6, 4.6, and 5.0\,eV for Pd, Fe, and Co ions, respectively~\cite{CALDERON2015233,Ostlin,hong}. 

In DFT, three corrective terms modify the Hamiltonian: the leading-order relativistic correction to the kinetic energy, the correction due to spin-orbit coupling (SOC), and the Darwin term, which arises from the quantum fluctuating motion or zitterbewegung of the electron. Scalar-relativistic (SR) calculations include the relativistic kinetic energy correction and the Darwin term. Fully relativistic calculations (SR + SOC) also incorporate SOC, which is essential to predict the magnetic anisotropy energy (MAE) because it drives the magnetocrystalline anisotropy contribution (MCA)~\cite{Blanco_2019}. 

In our approach, for the bulk and thin films of mono-component systems of Fe and Co, we first performed full optimization within SR calculations, including cell shape, volume, and ion positions. In contrast, for disordered systems such as low-impurity Pd-based alloys, full optimization was performed only to the pure host matrix. The Fe/Co impurity was then introduced by substituting a Pd atom without subsequent relaxation. This algorithm preserves the initial crystal symmetry and allows us to isolate symmetry-related contributions to the MAE. A similar procedure was used in Ref.\,\cite{lusakowski2017ab}. Indeed, we tested both relaxed and unrelaxed alloy cells and found that relaxation led to unreliable results, including a broad variation in MAE values and inconsistency in easy axis directions. These findings highlight the critical role of local symmetry in obtaining reliable MAE behavior. Distortions from relaxation in small supercells can hide real physical trends, while keeping the lattice fixed helps to better reveal the true anisotropy behavior in dilute metallic alloys.

Besides, for alloys in both bulk and film geometries, the averaging over three configurations was performed for the results' reproducibility. Various configurations correspond to the substitution of magnetic impurity with different Pd atoms.

The total energy for a given magnetization direction was calculated using SOC and applied self-consistently (SC) to valence electrons. This means that the electronic density and wave functions are iteratively updated until they converge with the fully incorporated SOC. We used this SC-method for all considered geometries since it offers accurate and reliable determination of total energies than the common non-SCF method reported in the literature.

The bulk structures of the alloy consist of a filled face-centered cubic (fcc) host matrix formed by Pd atoms with impurity atoms substituting for octahedrally coordinated sites. The initial Pd unit cell was fully relaxed and then used to construct the supercells. The unit cell of the alloy was contracted as 2$\times$2$\times$2 bulk unit cells of Pd with a metal substituting for one of the Pd atoms. This unit cell with chemical formula Pd$_{31}$\textit{Me} (\textit{Me}=Co, Fe) corresponds to the impurity concentration of approximately 3\,at.\%.

\section{Results}
\subsection{MAE calculations for bulk,  monolayers, and thin-films of Fe and Co}  
\label{bulk_monol} 
\subsubsection{Bulk Fe and Co}
In the first stage, we present the results for the bulk magnetic metals of Fe and Co to provide a methodology test and detailed analysis of the resulting electronic and magnetic features. Basic methodological parameters related to structure optimization and electronic structure predictions were thoroughly tested in our previous research~\cite{our_alloys_paper}, revealing a good coincidence with the available experimental structural parameters.

\begin{figure}[!h]
\centering
\includegraphics[width=\linewidth]{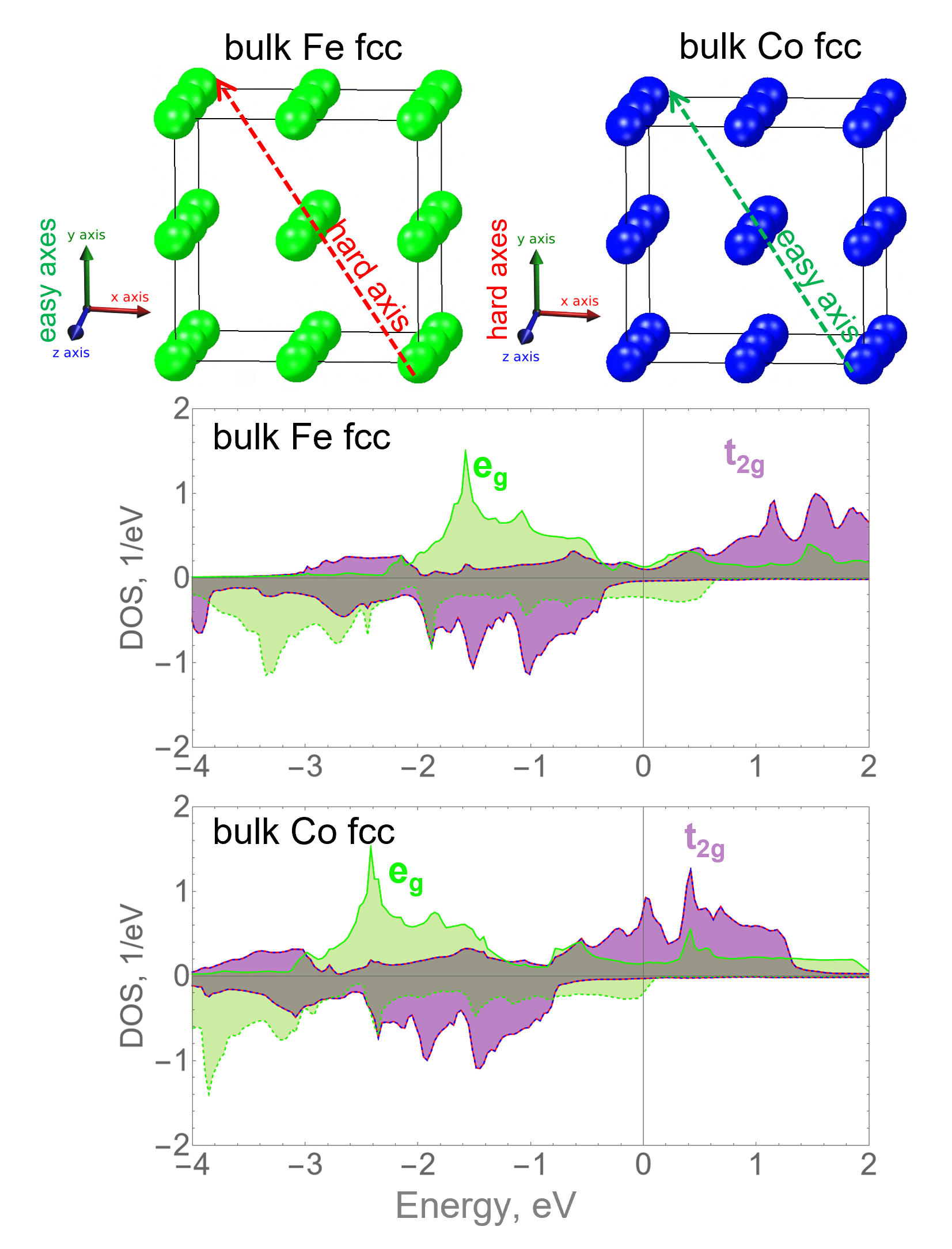}
\caption{Bulk bcc Fe (green balls) and Co (blue balls) in the unit cell representation with denoted diagonal [111] direction and corresponding spin- and orbital-resolved density of states.}
\label{fig_bulk}
\end{figure}

After a good relaxation of the structures, SOC calculations were conducted within the SCF scheme and high \textbf{k}-mesh. The directions of the easy and hard axes of magnetization were determined (as presented in Fig.\,\ref{fig_bulk}), and the magnetic anisotropy energies were calculated according to the equation: 
\begin{equation}
    MAE = E_{[001]} - E_{[111]}.
    \label{mae_bulk}
\end{equation}

The MAE values were calculated per Fe/Co atom and collected in Table\,\ref{tab_mae}, along with the available reference values from experiments and previous calculations. 
\begin{table*}[!ht]
\caption{Collection of anisotropy types in bulk, monolayers, and ultra-thin-films (four monolayers) along with corresponding MAE values calculated per Fe/Co atom in denoted units. The values from Ref.~\cite{li1990magnetic} are given for the Fe monolayer on Ag, Pd, and Au substrates, respectively.}
\centering
\begin{tabular}{p{5cm}|lp{3cm}l}
\textbf{Bulk }           & fcc Fe  &  bcc Fe   & fcc Co \\ \hline
Anisotropy   type        & hard [111] &  & easy [111]  \\
MAE, $\mu$eV/atom        &   -709.2    & -42.5 & 19.5 \\
Exp. MAE,  $\mu$eV/atom  &     &   -1.4~\cite{tung1982magnetic} & 1.8~\cite{trygg1995total}   \\ \hline    
\textbf{Monolayer}  \\ \hline
Anisotropy   type (easy axis)  & out-of-plane & & in-plane \\
MAE, meV/atom                       & -0.23        &  & 1.08 \\
Calc. MAE from Refs,  meV/atom   &   \begin{tabular}[c]{@{}l@{}} -0.06, -0.35, -0.57~\cite{li1990magnetic},\\ -1.95~\cite{lorenz1996magnetic} \end{tabular}     &  &    \\ \hline
\textbf{Thin-film}  \\ \hline
Anisotropy   type (easy axis)  & in-plane & & in-plane \\
MAE, eV/atom                   & 0.19    &  & 0.07
\label{tab_mae}              
\end{tabular}
\end{table*}
For Co, which was taken in an fcc structure, the easy axis of magnetization has the direction [111]. The axes \textit{Ox}, \textit{Oy}, and \textit{Oz} were found to be the hard axes.

The situation is the opposite for iron (Fe). Fe in the present research was considered in the fcc phase (to compare with the same phase of Co and also to the Pd-alloys, where Pd has the fcc cell) and in a more general bcc phase for comparison. The fcc structure for iron is a metastable phase and exhibits ferromagnetism. However, the magnetic ordering in fcc Fe is generally weaker compared to the more stable bcc Fe~\cite{tung1982magnetic}. As for the magnetic anisotropy, the hard axis of magnetization for both phases of Fe was found to be in the [111] direction, and \textit{Ox}, \textit{Oy}, and \textit{Oz} are the easy axes, respectively. The directions of magnetic anisotropy in bulk structures are shown in Fig.\,\ref{fig_bulk}, where the unit cells are presented along with the density of states (DOS) with orbital resolution.  

MAE of bcc Fe shows a smaller value compared to that of the fcc phase. A significant discrepancy is observed between the calculated and experimental values. However, the literature review indicates that theoretical numbers are strongly sensitive to the type of calculation. For example, the first-principles study by Razee and coauthors has shown a strong variation of MAE values~\cite{razee1997first}. Besides, the comparison with different \textbf{k}-points and computational schemes was performed within the present research, and the results quantitatively vary significantly. Consequently, only qualitative comparisons in terms of hard and easy axes are available here, which shows a reasonable agreement.

The difference between bulk fcc Fe and Co can also be analyzed by the electronic structure comparison of their 3\textit{d} states. As seen from the DOS plots in Fig.\,\ref{fig_bulk}, the key distinction lies in the distribution of their $e_g$ and $t_{2g}$ orbitals near the Fermi level and their difference in spin-up and spin-down states, which influence the magnetic properties and also can affect the magnetic anisotropy through the spin-orbit coupling. For both Fe and Co, the $t_{2g}$ states lie slightly below the $e_g$ states, in agreement with \textit{d}-orbital splitting in an octahedral environment. Fo Co $t_{2g}$ states form a pronounced peak at the Fermi level, while the contribution from the $e_g$ states is less, and DOS for spin up and down are equal. In contrast, for Fe, while the $t_{2g}$ states are still present at the Fermi level, their dominance is significantly reduced compared to Co. At the same time, the filling of spin-up and spin-down components is symmetrical as well. The presented DOS analysis does not allow directly make a direct conclusion about the easy axis direction. Indeed, as was pointed out in Ref.~\cite{razee1997first}, even states that being far removed from the Fermi level can also contribute to the MAE, and a slight shift in the Fermi level can alter the magnitude of MAE as well as the direction of the magnetic easy axis.

\subsubsection{Fe and Co monolayers}
\label{monolayers}
Further, similar calculations were performed for monolayers of pure magnetic compounds. Due to the structural symmetry of the monolayer systems, the directions [100], [010], and [110] were found to be energetically equal, so we were interested only in the difference of magnetic anisotropy in the plane or out of the monolayer plane. The MAE for monolayers was calculated as a difference in the system's energies when spins are oriented along \textit{z}-direction [001] and the in-plane direction [100] or [010], which are equal due to symmetry: 
\begin{equation}
    MAE = E_{[001]} - E_{[100]}.
    \label{mae_monolayer}
\end{equation}

\begin{figure}[!h]
\centering
\includegraphics[width=\linewidth]{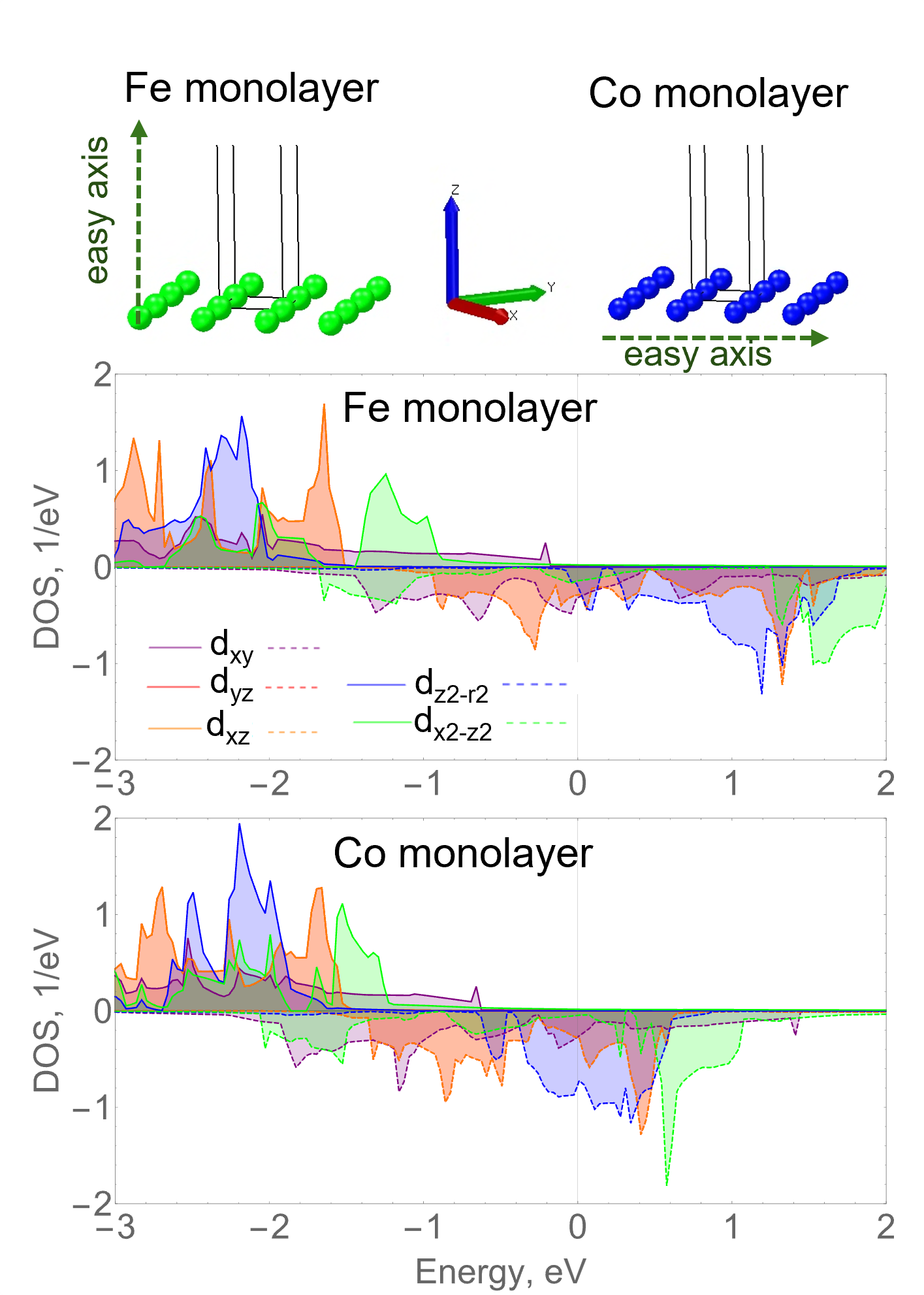}
\caption{Monolayers of Fe and Co in the unit cell representation with denoted easy axis direction and corresponding spin- and orbital-resolved density of states.}
\label{fig_monolayers}
\end{figure}

It was found that for the Fe monolayer, the easy axis of magnetization is perpendicular to the plane, and for Co, oppositely, the easy axis lies in the plane of the layer (Fig.\,\ref{fig_monolayers}). 
The obtained funding agrees with the experimental research of Kief et al.~\cite{10.1063/1.352695} for a Fe bilayer on Cu substrate, where the out-of-plane anisotropy was observed. Furthermore, the qualitative agreement was reached for the Fe monolayer calculations mentioned in Ref.~\cite{li1990magnetic} performed for a monolayer on Au, Ag, and Pd (001) substrates. Besides, the same types of anisotropy (perpendicular for Fe and in-plane for Co) were found in Ref.~\cite{lehnert2010magnetic} for HCP phases on Rh(111) and Pt(111) substrates.

The electronic structure was further analyzed to reveal the possible origin of magnetic properties in selected systems. In comparison with bulk DOSes, the distribution of states is much more complicated, which arises due to cubic symmetry breaking. For both Fe and Co monolayers, only $d_{xz}$ and $d_{yz}$ remain nondegenerate due to symmetry. In addition, both monolayers demonstrate half-metallic behavior with Fermi level contribution for spin-down components only. For the Fe monolayer, the most pronounced orbitals near the Fermi level were found to be $d_{xy}$, a slightly smaller is $d_{xz}$ and $d_{yz}$, which are all t$_{2g}$ orbitals.

For the Co monolayer, there are more states near the Fermi level compared to Fe.  For Co, the main contribution at the Fermi level is from the spin-down component of $d_{z^2-r^2}$ (twice larger than any other orbital), which is elongated along the \textit{z}-direction. That orbitals' distribution possibly might lead to the occurrence of a hard axis in the out-of-plane direction. At the same time, the in-plane $d_{x^2-y^2}$ orbital has the lowest contribution, making the monolayer plane an easy direction.

\subsubsection{Fe and Co thin films}
\label{films}
\begin{figure}[!h]
\centering
\includegraphics[width=\linewidth]{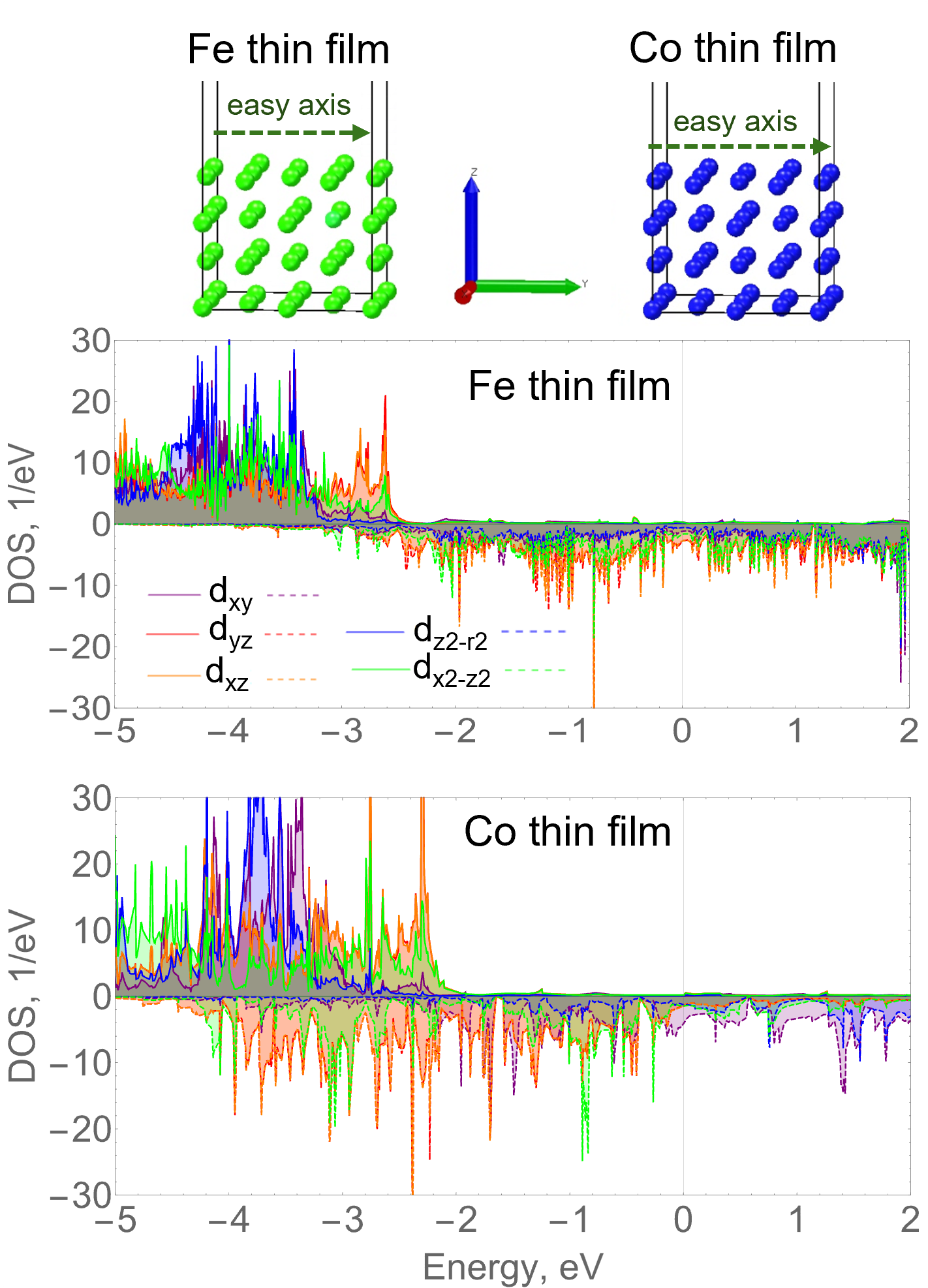} \\
\caption{Four-layered structures of Fe and Co in the unit cell representation with denoted easy axis direction and corresponding spin- and orbital-resolved density of states.} 
\label{fig:films}
\end{figure}

To follow the changes in the structure and magnetic properties relation, the four-layered slabs (as presented in Fig.\,\ref{fig:films}) were constructed in the same way as for the monolayers. 
The strength of anisotropy, defined in MAE values, for the investigated systems strongly depends on the geometry and the difference of three orders of magnitude, as seen from Table\,\ref{tab_mae}.
Interestingly, Fe changes the anisotropy type from out-of-plane for monolayer geometry to in-plane for the four-layered structure, as also shown in Fig.\,\ref{fig:films}. We have also evaluated that this thickness is critical, and below it, the preferred orientation is out-of-plane.

While comparing the DOS plots from Fig.\,\ref{fig_monolayers} of monolayer with DOSes from Fig.\,\ref{fig:films}, one can see that the number of states significantly increases, making the analysis complicated. In the case of four-layered Fe, at the Fermi level $d_{xz}$ and d$_{yz}$ have the most contribution, which are not in plane, as was the case for the monolayer, with the most differences between $d_{xy}$, $d_{xz}$, and $d_{yz}$. 

For Co in the thin film geometry (Fig.\,\ref{fig:films}), the easy axis does not change direction, preserving the magnetization in the plane. Even though the largest contribution at the Fermi-level was found for $d_{xy}$, all other orbitals also contribute equally.

\subsection{MAE calculations for Pd$_{0.97}$Co$_{0.03}$ alloy in bulk and ultra-thin film geometries} 
\label{results_alloys}
\begin{figure}[!h]
\centering
\includegraphics[width=\linewidth]{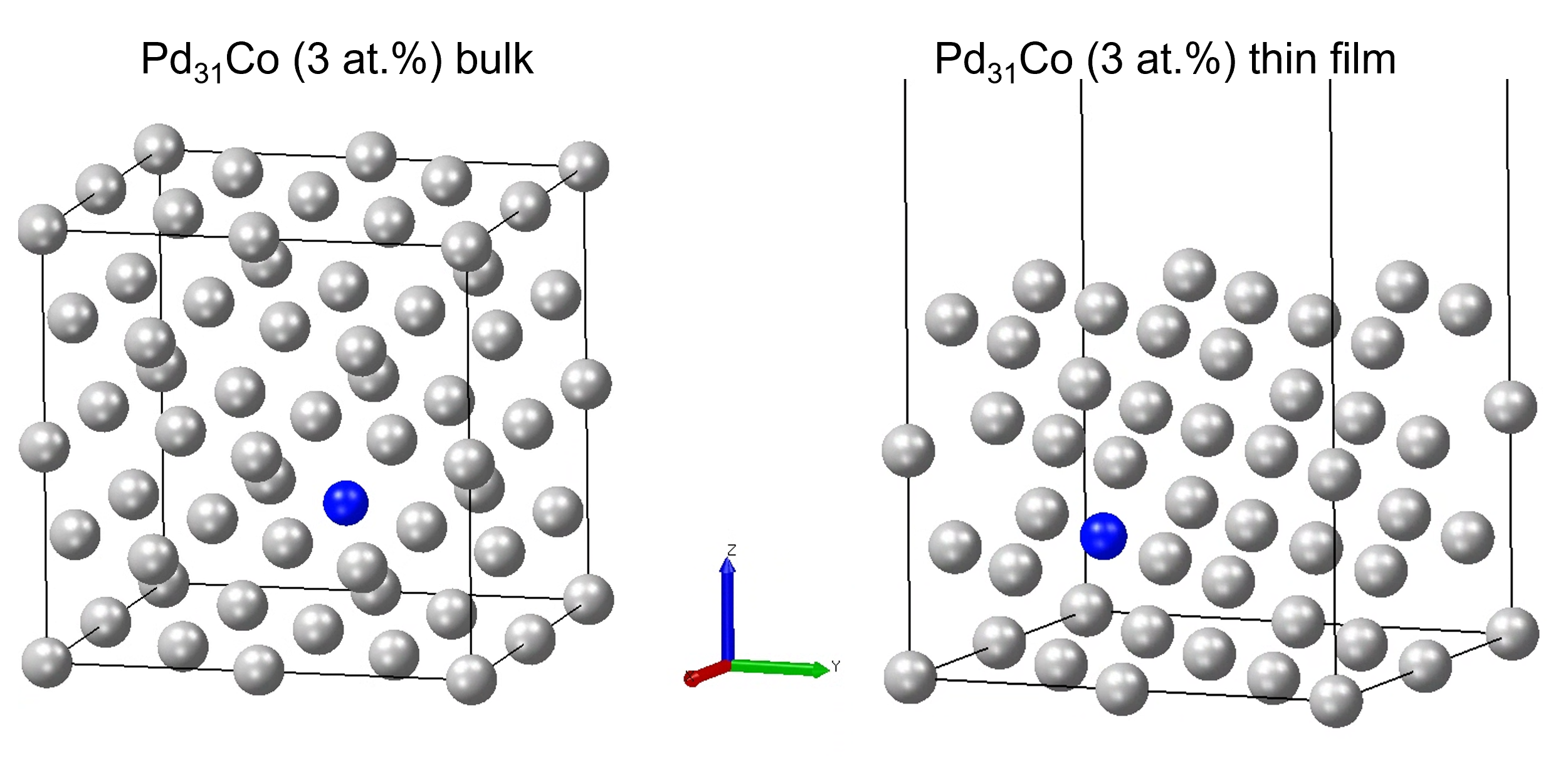} \\
\caption{Unit cells used for calculations for alloys (Pd$_{31}$Co 3\,at.\%) in bulk and ultra-thin-film geometries.} 
\label{fig_alloy}
\end{figure}

At the next stage, the bulk and ultra-thin film unit cell construction was performed for Pd$_{0.97}$Co$_{0.03}$ (or Pd$_{31}$Co) alloy, which corresponds to the impurity concentration of \textit{x} = 3\,at.\%. The corresponding unit cells are presented in Fig.\,\ref{fig_alloy}.
The choice of concentration was based on the need to balance computational efficiency and the clarity of the results. Lower concentrations necessitate larger computational cells, complicating non-collinear calculations. Furthermore, selecting the lowest feasible concentration maximizes the observable effects, enhancing the validity of our findings, as highlighted in Sec.\,\ref{computations}.

As was already mentioned, the calculation of MAE involves the calculation of the energy difference for the system with magnetization aligned along various crystallographic axes. For alloys, we examined the high-symmetry axes of the crystal structure, including [111], which corresponds to the body diagonal and intersects with in-plane directions such as [110], [100], and [010], as well as the out-of-plane component [001]. We also considered the arbitrary direction [112], which is particularly relevant for film geometries, as it helps identify potential strain effects in the film. Consequently, the MAE for the alloys was calculated as a difference in the system's energies when spins are oriented along [$uvw$] and the most favorable direction (easy axis): 
\begin{equation}
    MAE = E_{[uvw]} - E_{easy}.
    \label{mae_alloy}
\end{equation}

\begin{figure}[!h]
\centering
\includegraphics[width=0.9\linewidth]{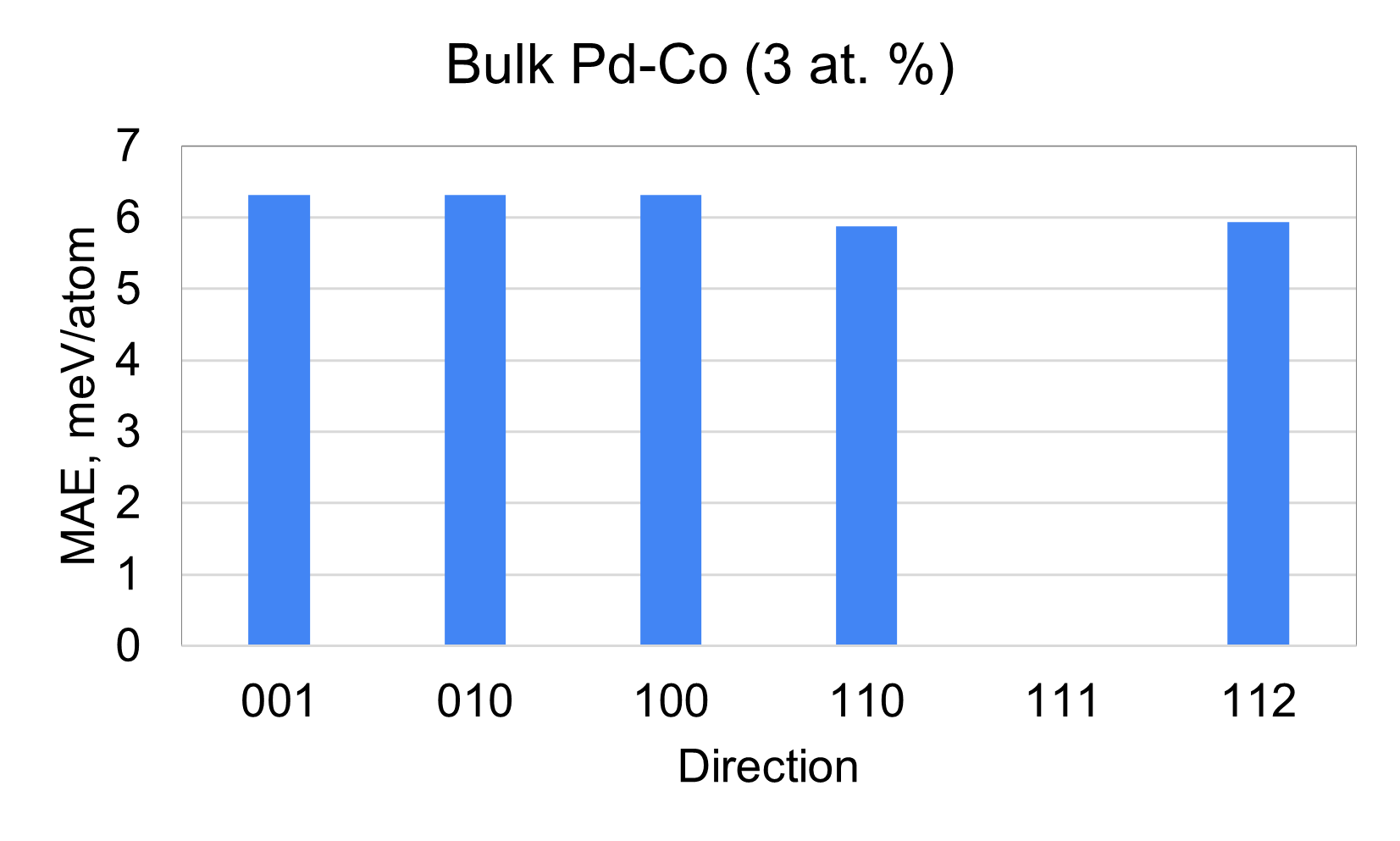} \\
\includegraphics[width=0.9\linewidth]{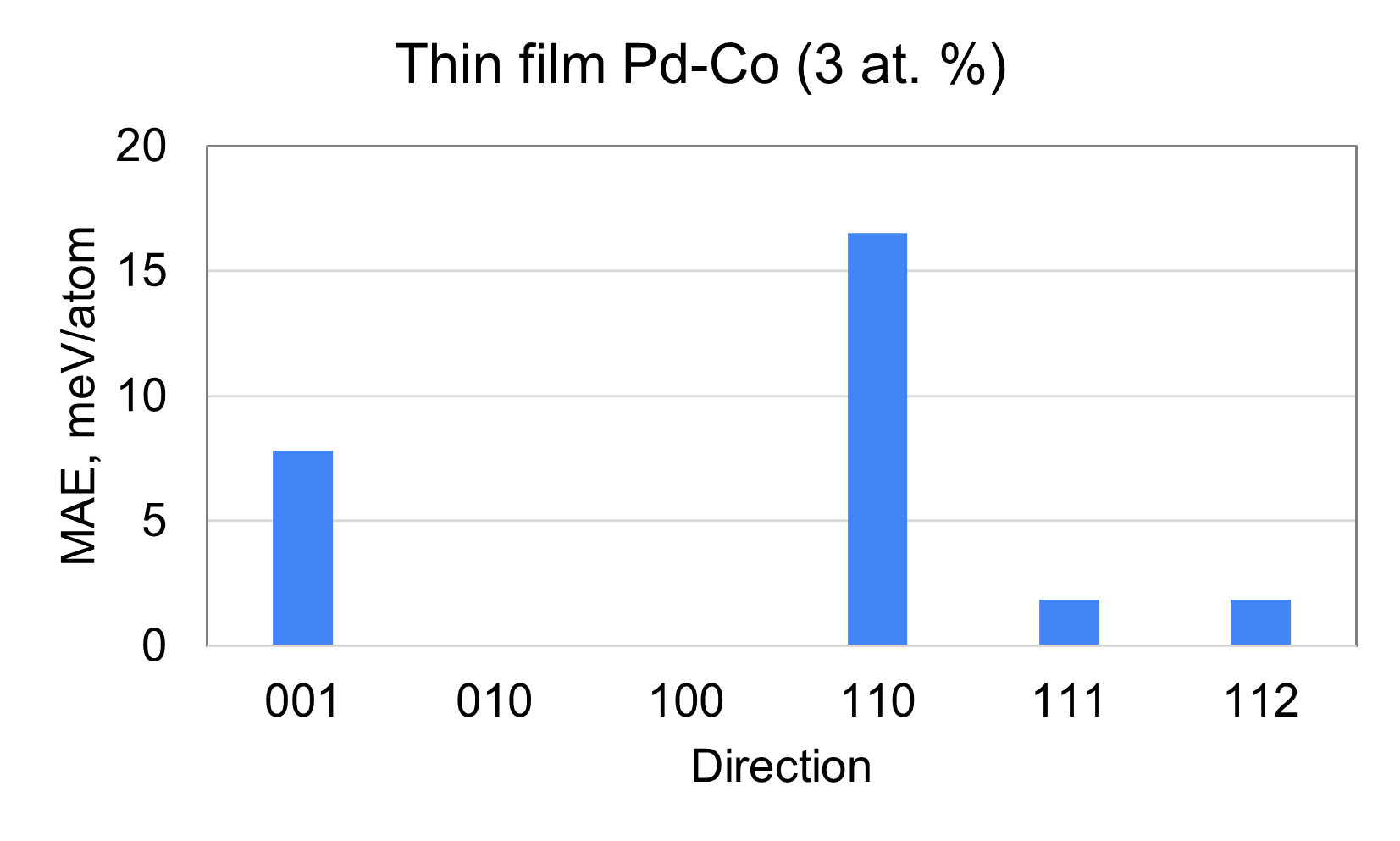} \\
\caption{MAE calculated for Pd$_{0.97}$Co$_{0.03}$ in bulk and thin film geometries averaged over three configurations.} 
\label{fig:alloy_gist}
\end{figure}

The calculated MAEs for CoPd alloy are reflected in the histograms of Fig.\,\ref{fig:alloy_gist}, where each bar corresponds to the MAE energy calculated for a specific direction of a cell according to Eq.\,\ref{mae_alloy}. 
In the case of bulk Pd$_{31}$Co (Pd$_{0.97}$Co$_{0.03}$) (Fig.\,\ref{fig:alloy_gist}), a single easy axis is observed along the [111] direction, while all other directions exhibit an average MAE of 6.15\,meV/atom. This behavior coincides with the bulk Co, where the body diagonal is also the easy axis (Fig.\,\ref{fig_bulk}). The observed finding also agrees well with experimental evidence. For example, Ref.~\cite{bagguley1974resonance} reports that in bulk Pd-Co with 1.0\,at.\%, and 1.5\,at.\%, the [111] direction is associated with an easy axis.

In the geometry of an ultra-thin Pd$_{31}$Co film (Fig.\,\ref{fig:alloy_gist}), the hard axis was found along the in-plane diagonal [110] direction, whereas the other in-plane [100] and [010] directions became easy axes. This trend coincides with the behavior of pure Co in monolayer and thin-film structures. The highest MAE of 16.53\,meV/atom was found along the [110] direction, while the out-of-plane [001] direction was also hard, with a lower MAE of 7.81\,meV/atom.  

The observed behavior of MCA in the Pd-Co alloy highlights that even a small Co concentration is sufficient to induce magnetic anisotropy trends that closely resemble those of bulk Co and pure Co thin films.
\begin{figure*}[ht!]
\centering
\includegraphics[width=0.9\linewidth]{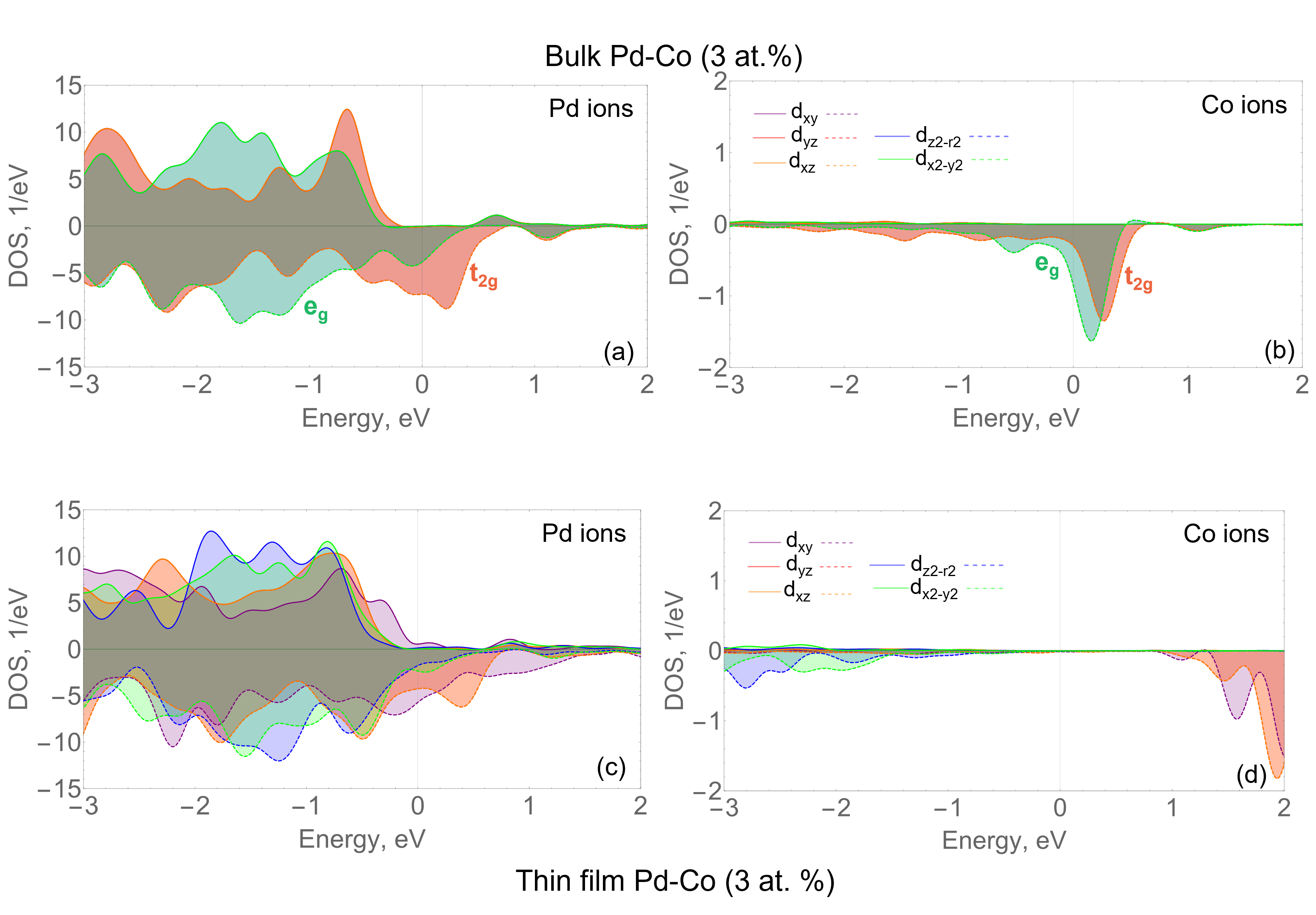}
\caption{Spin- and orbital-resolved DOS for Pd$_{0.97}$Co$_{0.03}$ alloy plotted separately for Co and Pd: (a) and (b) in bulk geometry, (c) and (d) in ultra-thin film, corresponding to unit cells depicted in Fig.\,\ref{fig_alloy}.}
\label{fig_dos_alloy}
\end{figure*}
The possible reasons for the observed anisotropic behavior in Pd-based alloys might be linked to the electronic structure. The atom and orbital resolved DOSes for Pd and Co atoms of Pd$_{0.97}$Co$_{0.03}$ alloy are presented in Fig.\,\ref{fig_dos_alloy}.

As seen from DOS plots of Pd-Co alloy in bulk geometry Fig.\,\ref{fig_dos_alloy} a and b), both $t_{2g}$ and $e_g$ orbitals are degenerated due to cubic symmetry. At the Fermi level, it is seen that the system is a semi-metal and the conductivity occurs via spin-down electrons. The biggest contribution at the Fermi level comes from  $t_{2g}$ orbitals as in the bulk Co (Fig.\,\ref{fig_bulk}). As for the Co atom contribution, the largest contribution at the Fermi level was found for the $e_g$ orbital, but the DOS is smaller than for Pd.

In comparison, in the film geometry, the spin-splitting is more complicated due to lower symmetry. It also possesses semi-metallic character with the main contribution at the Fermi-level from the spin-down component as presented in Fig.\,\ref{fig_dos_alloy}\,c. In Pd ions, the biggest density of states at the Fermi level was found for the in-plane $d_{xy}$ and slightly smaller for $d_{yz}$ and $d_{xz}$, whereas other $e_{g}$ orbitals are less and equally filled. Besides, the Co atom levels are located away from the Fermi level, not contributing significantly to the electronic and magnetic properties.

\begin{figure}[!h]
\centering
\includegraphics[width=0.95\linewidth]{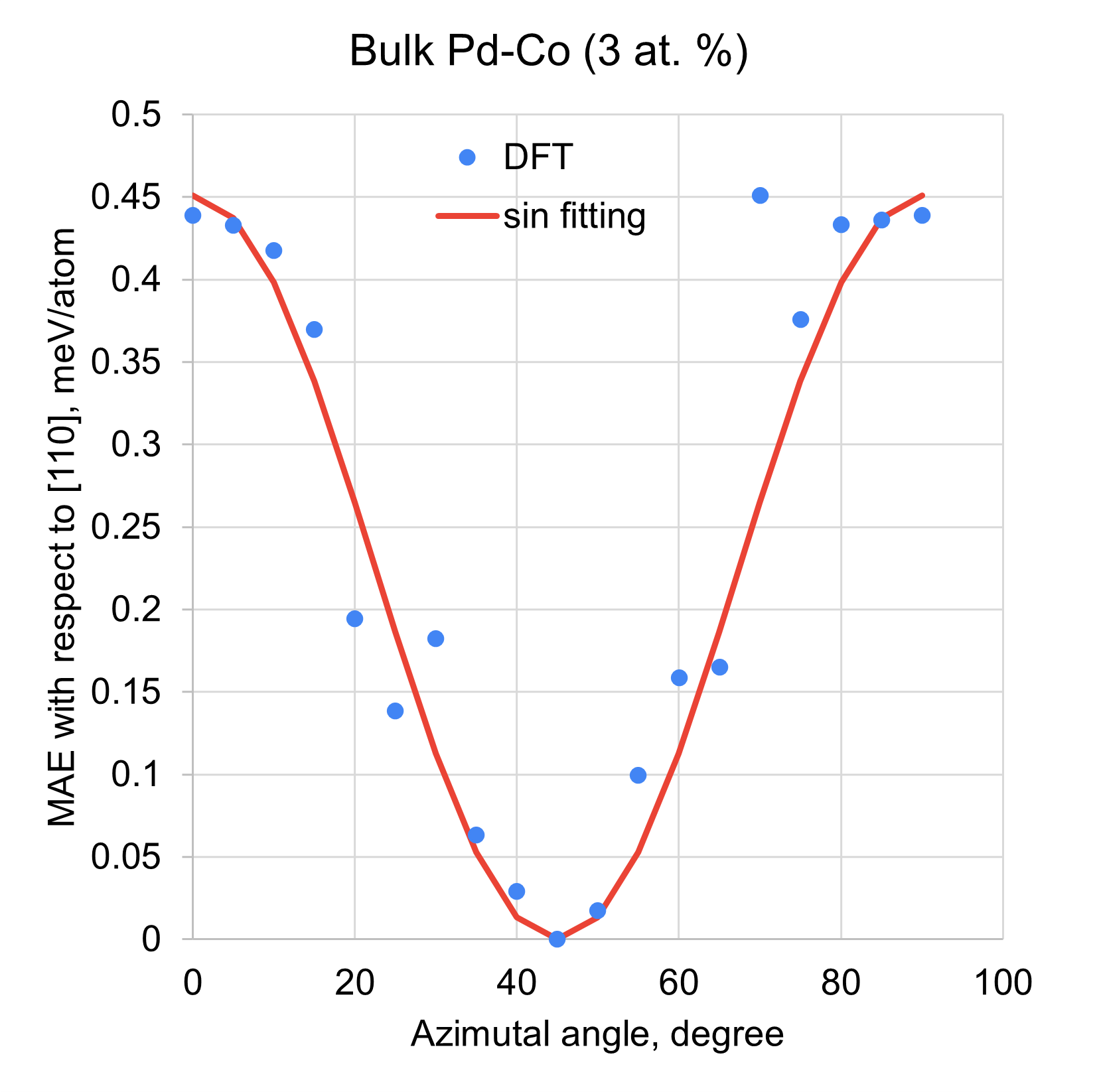} \\
\caption{Variation of MAE in \textit{xy}-plane calculated for Pd$_{0.97}$Co$_{0.03}$ in bulk geometry (blue dots) along with sine-squared fitting (solid red line).} 
\label{fig:alloy_plane}
\end{figure}

In both geometries, there is no isotropic distribution of MAE for \textit{xy}-plane directions. Thus the calculations of MAE for Pd$_{0.97}$Co$_{0.03}$ was separately explored according to Eq.\,\ref{mae_alloy}, with spins varying in \textit{xy}-plane, i.e., having [$uv$0] direction, which corresponds to the change of azimuthal angle only when polar is fixed at $\pi$/2. The results for the bulk geometry are presented in Fig.\,\ref{fig:alloy_plane}.
MAE exhibited a smooth, sine-squared dependence on the azimuthal angle within the \textit{xy}-plane, consistent with expectations for systems with high cubic symmetry. As seen, the easy [111] direction of magnetization is projected onto the \textit{xy}-plane and forms an angle of 45 degrees with the \textit{x}-axis, repeating every 90 degrees ($\pi$/4+$\pi$*n/2, n = 0--1). From the fitting with the sine squared law, the anisotropy constant was extracted and equals 0.45\,meV/atom.

\begin{figure}[!h]
\centering
\includegraphics[width=0.95\linewidth]{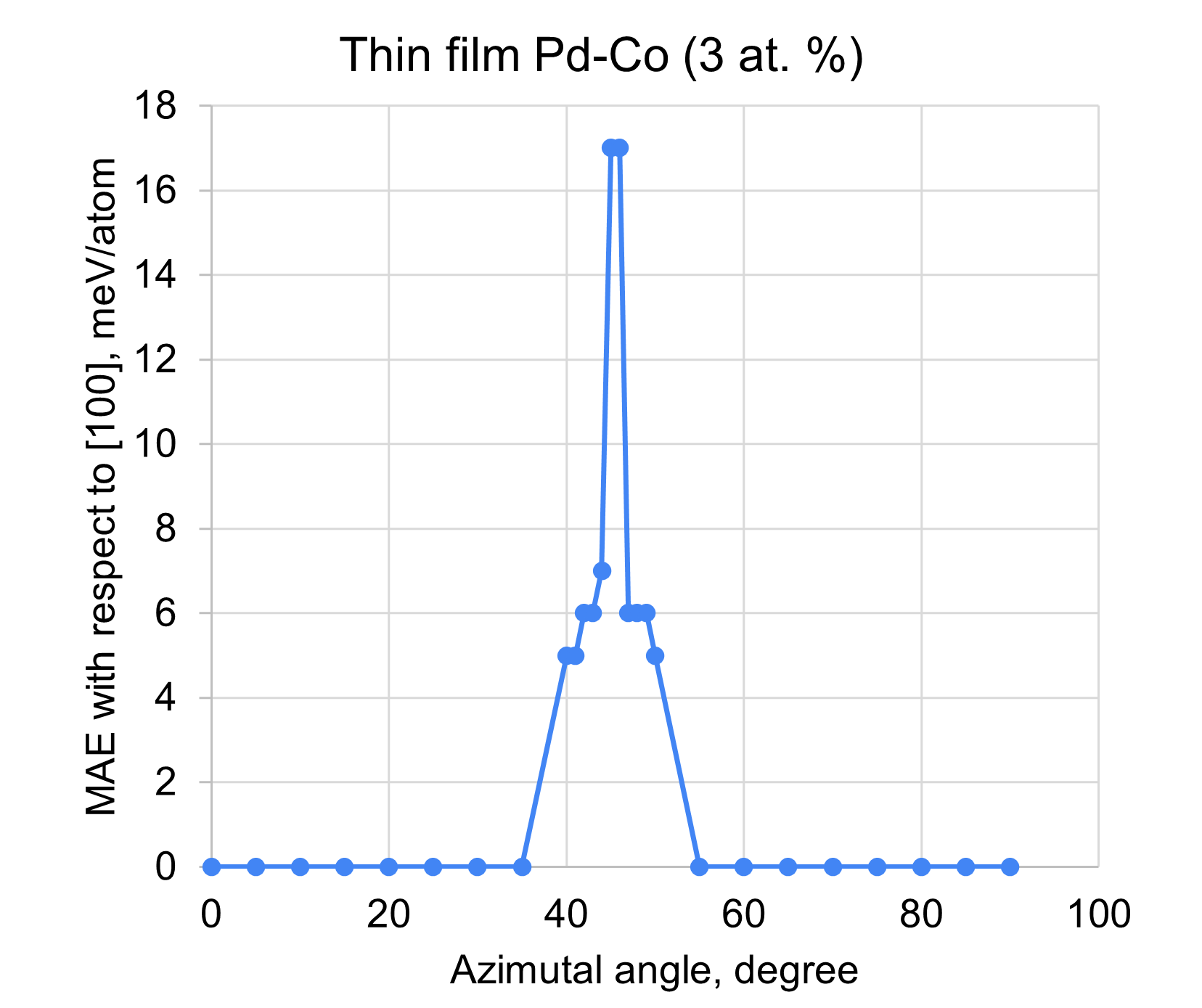} \\
\caption{Variation of MAE in \textit{xy}-plane calculated for Pd$_{0.97}$Co$_{0.03}$ in thin film geometry (blue dots connected with solid line for better visual perception).} 
\label{fig:alloy_plane_film}
\end{figure}

In the restricted geometry, in the case of ultra-thin Pd-Co films, the hard [110] magnetization direction in the \textit{xy}-plane forms an angle of 45 degrees with the \textit{x}-axis, also repeating every 90 degrees ($\pi$/4+$\pi$*n/2, n = 0--1) (Fig.\,\ref{fig:alloy_plane_film}. The azimuthal angle (variation) does not represent a sinusoidal law, and the hard direction forms a very sharp and intense peak.
This finding is required to be checked by experimental techniques and will be a subject of our further experimental investigations for Pd-Co system films in the 1-10 at.\,\% impurity concentration range.

\subsection{Pd$_{0.97}$Fe$_{0.03}$ alloy}
Lastly, we would like to address the case of Pd-Fe alloy. In contrast to the Pd-Co case discussed above, the bulk Pd-Fe alloy with the same 3\,\% concentration was found to be isotropic (all three considered configurations). We expect that the situation might change with other higher Fe concentrations, since indirect interactions can stabilize anisotropy when impurity concentration increases. Indeed, in experiments, for example, the investigation of Bagguley and Robertson for bulk Pd-Fe has revealed isotropic magnetic behavior of the alloy with 0.25\,at.\%, whereas at 1.7\,at.\% the [111] direction was found as an easy axis~\cite{bagguley1974resonance}. 

In contrast to the MAE in Pd-Co alloys, the calculations for the thin Pd-Fe film did not show consistent results across different configurations, despite yielding significant magnitudes of magnetic anisotropy energy.  The underlying cause of this inconsistency in Fe's behavior remains unclear and warrants further investigation. However, the possible explanation might be the following. In dilute Pd-Fe, Fe atoms remain isolated, and their magnetic moments are highly sensitive to the local environment~\cite{piyanzina_crystals}. Unlike Co, which has a stable moment in Pd, Fe strongly interacts with the surrounding Pd states, leading to variable hybridization and local distortions. This causes significant fluctuations in magnetic anisotropy depending on Fe’s position. 

We also estimated the shape anisotropy for the thin film geometry of Pd-Co and Pd-Fe using the following procedure. For a thin film, the shape anisotropy energy per atom is approximately:
\begin{equation}
    E_{shape} = 1/2 \mu_0 M_s^2(N_z-N_x)\approx1/2 \mu_0 M_s^2,
    \label{mae_shape}
\end{equation}
where $\mu_0$ is the permeability of free space, $M_s$ is saturation magnetization, and demagnetization coefficients $N_z\approx1$ and $N_x=N_y=0$ for the very thin films, where the shape anisotropy favors in-plane magnetization. We obtained a value of $1.45\times10^{-3}$\,meV/atom for Pd$_{0.97}$Co$_{0.03}$, which is three orders of magnitude less than magnetocrystalline anisotropy, estimated as 16.53\,meV/atom. The estimation of the shape anisotropy energies revealed a slightly higher value for the Pd-Fe alloy than for the Pd-Co case, being $3.21\times10^{-3}$\,meV/atom.

\section{Conclusion}
\label{concl}
In the present work, we have demonstrated the potential of DFT methods in describing magnetocrystalline anisotropy properties for bulk Fe and Co, the monolayers and thin films of these magnetic metals, and low-impurity Pd-Co and Pd-Fe alloys. By performing self-consistent noncollinear calculations, we revealed that for the bulk Fe and Co in fcc phases, the diagonal [111] axis is a hard and easy one, respectively.

For monolayers, the opposite characteristics are also present. The Co monolayer has an easy axis in the in-plane direction, whereas the Fe possesses a significant out-of-plane orientation in consistent with the previous calculations and experiments. While increasing the thickness of the slabs, the Co film preserves the in-plane anisotropy, whereas Fe changes the direction of the easy magnetization axis to the in-plane.

Lastly, for the low-impurity alloys, we confirmed that the direction of the easy axis depends significantly on the geometry of the sample. In particular, in the bulk Pd$_{0.97}$Co$_{0.03}$ alloy, the easy magnetization direction coincides with the diagonal [111] axis. In contrast, in the film geometry, the [110] and less [001] directions become most unfavorable, and the film demonstrated in-plane anisotropy. Finally, the MAE in bulk alloy exhibits a smooth, classical for cubic symmetry sine-squared dependence on the azimuthal angle, whereas the film has a sharp and intense hard direction.

In summary, our results provide a comprehensive overview of the anisotropy properties of Fe and Co in various geometries and offer important new insights into the mechanisms of magnetic anisotropy in dilute Pd-Co alloys, which are challenging to study experimentally.

\begin{acknowledgments}
The work of I. Piyanzina (Gumarova) was supported by the grant 24PostDoc/2-2F006 by the Higher Education and Science Committee of Armenia. Computational resources were provided by the Armenian National Supercomputing Center (ANSCC).
\end{acknowledgments}

\bibliography{apssamp}% Produces the bibliography via BibTeX.
%\bibliography{cas-refs}
\bibliographystyle{plain}
\end{document}